\documentclass[aip,rsi,reprint]{revtex4-1}
\usepackage{amsmath,amssymb}
\usepackage{iftex}
\usepackage{pifont}
\usepackage{textcomp}
\usepackage{mathcomp}
\usepackage{float}
\usepackage{gensymb}
\usepackage{tikz}
\usetikzlibrary{calc}
\usetikzlibrary{positioning}
\usetikzlibrary{arrows.meta}
\usepackage{graphicx}
\usepackage{dcolumn}
\usepackage{bm}
\usepackage[utf8]{inputenc}
\usepackage[T1]{fontenc}
\usepackage{mathptmx}
\usepackage{etoolbox}

\makeatletter
\def\@email#1#2{%
 \endgroup
 \patchcmd{\titleblock@produce}
  {\frontmatter@RRAPformat}
  {\frontmatter@RRAPformat{\produce@RRAP{*#1\href{mailto:#2}{#2}}}\frontmatter@RRAPformat}
  {}{}
}
\makeatother
\begin{document}

\title[]{High-resolution computed tomography of two-dimensional beam profile using dual-axis rotating wire}

\author{R.~Ota}
\altaffiliation{Department of Physics, Rikkyo University, 171-8501 Tokyo, Japan.}

\author{N.~Nakajima}%
\altaffiliation{Department of Physics, Rikkyo University, 171-8501 Tokyo, Japan.}

\author{R.~Takemasa}%
\altaffiliation{Department of Physics, Rikkyo University, 171-8501 Tokyo, Japan.}

\author{H.~Tamaru}%
\altaffiliation{Department of Physics, Rikkyo University, 171-8501 Tokyo, Japan.}

\author{Y.~Shiina}%
\altaffiliation{Department of Physics, Rikkyo University, 171-8501 Tokyo, Japan.}

\author{Y.~Nakano$^\ast$}%
\altaffiliation{Department of Physics, Rikkyo University, 171-8501 Tokyo, Japan.}
\email{nakano@rikkyo.ac.jp}

\date{\today}

\begin{abstract}
The use of a wire probe is a robust method for beam profile measurement, but it can only provide a 1D projection of the beam profile.
In this study, we developed a novel method for measuring a beam projected from a 360$\degree$ angle by a dual-axis rotation of a wire and obtaining a complete 2D profile via image reconstruction. We conducted a proof-of-principle study using an Ar$^+$ ion beam and optimized the reconstruction algorithm. 
The experimental results showed that the use of the order subsets expectation maximization (OS-EM) algorithm is the most reasonable method, providing a highly accurate absolute 2D beam profile within a processing time on the millisecond scale. 
Furthermore, analysis of 2D profiles at different probing positions provided the beam direction and the phase-space distribution. 
This versatile method can be applied to various fields of particle beam technologies, such as particle therapy, semiconductor processing, and material analysis, as well as basic scientific research.
\end{abstract}

\maketitle

\section{Introduction}
Diagnosis of beam profiles is a vital technology in most of the particle-beam-based research, medical, and industrial applications. 
Beam profile monitors with thin wires are especially popular due to their durability and versatility.
In fixed-wire detectors such as wire chambers, beam profiles can be obtained by detecting the beam current or secondary current flowing into each of their equally spaced wires. 
Alternatively, a single wire is used to measure the current by scanning its position relative to the beam, thereby generating continuous beam intensity profiles.
Based on these principles, various combinations of wire shapes and scanning motions have been developed\cite{Seely2008,Harryman2017,Arutunian2021a,Igarashi2002,Stratakis2006}.
A critical advantage of the wire scanners is their extensive dynamic range of measurable beam intensities, which ranges from picoamperes to tens of Amperes. However, in principle, wires can only provide a 1D projection of the beam profile along the fixed array or the scanning direction of the wire. 

Tomographic technique allows for reconstruction of 2D beam profiles from a finite number of 1D projections taken from different directions.
Rotation of the beam in phase space by quadrupoles or solenoids allows tomographic reconstruction of the phase-space profile of the beam\cite{McKee1995,Xiang2009}.  
To obtain a real-space 2D beam profile, Xing \textit{et al}.\cite{Xing2018} used a multi-wire scanner and reconstructed the 2D profile of high-current proton beam with reasonable accuracy. 
Moretto-Capelle \textit{et al}.\cite{Moretto-Capelle2023} used a single-wire scanner combining rotary and linear reciprocating motions to measure 1D projections and reconstruct the 2D profile of a 200~eV electron beam. 
These studies have demonstrated the potential advantages of wire scanners for 2D beam profiling. 
Despite the significant technical challenges remaining, the feasibility of this approach offers a promising future for beam diagnostics, potentially surpassing existing methods using 2D detectors.

The reconstructed 2D images in previous studies contained reconstruction artifacts, which are distortions or errors that do not represent the actual structures. 
Thus the data sampling methods and reconstruction algorithms should be improved urgently to suppress these artifacts and ensure the accuracy of the reconstructed beam profiles.
The position resolution of these methods should also be quantitatively evaluated. 
In addition, the intensity of the reconstructed 2D beam image should be converted into an absolute beam flux profile to facilitate the practical application of the proposed method as a beam diagnostic device.

The present study aimed to establish the optimal approach to accurate high-resolution absolute 2D beam profiling using scanning wire tomography. 
We designed a novel dual-axis wire scanning device and investigated image reconstruction accuracy using different sampling and reconstruction algorithms. 
We also show that in addition to the 2D beam profile, the phase-space distribution of the beam can also be measured.
In principle, this method can be implemented with ease by introducing a single axis of rotational motion from outside of the vacuum chamber, instilling a wide feasibility of the proposed method.

\section{Principle of dual-axis rotating-wire tomography}
\subsection{Measurement of 1D projections}

\begin{figure*}
\includegraphics[width=\linewidth]{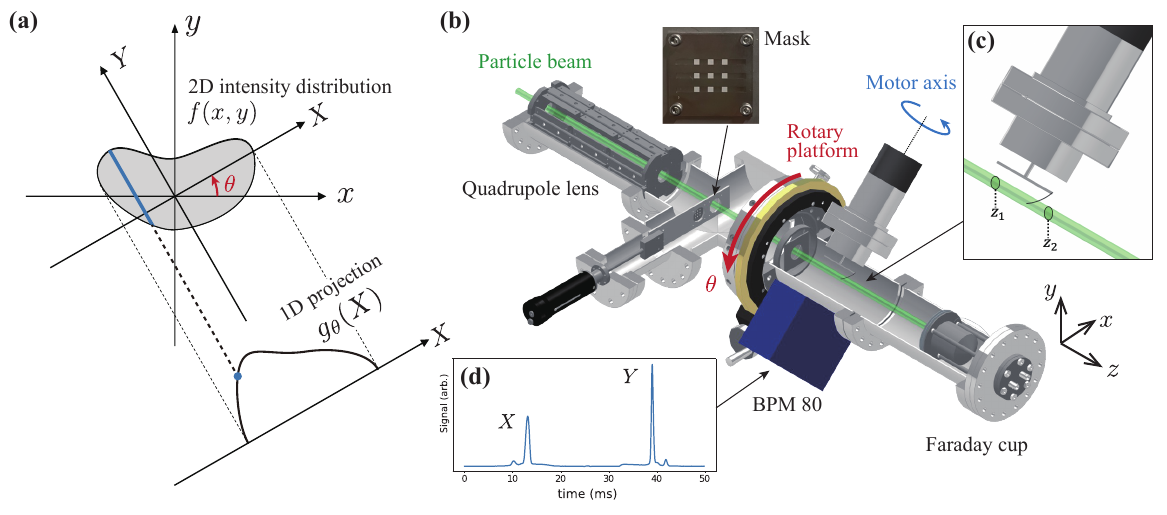}
\caption{
(a) Principle of measuring the 1D projection of a beam cross section using a scanning probe wire. In this study, the projection angle $\theta$ is changed by rotating the wire scanner itself with respect to the beamline. (b) Schematic illustration of the experimental setup for dual-axis rotating-wire tomography. The particle beam passes through the quadrupole electrostatic lens, the mask, and the wire scanner, which is mounted on a rotary platform. A photograph of the grid mask used in this study is shown. (c) Enlarged view of the rotating wire in the scanner. The wire crosses the beam at positions $z_1$ and $z_2$ at each turn. (d) Example of the current signal observed through a oscilloscope, where the beam current is a function of the rotating-wire position.  
}
\label{fig:setup}
\end{figure*}

Assume a beam propagates along the $z$~axis with a cross-sectional intensity distribution $f(x,y)$. 
When a wire is inserted into the beamline, it detects the amount of beam hitting the wire.
By scanning a straight wire aligned on an arbitrary $Y$ axis along the perpendicular direction $X$ across the $xy$ plane, one can measure the 1D projection of the beam profile $g_\theta(X)$. Here, the projection angle $\theta$ denotes the angle between the $x$-axis and the scanning direction $X$, as illustrated in Fig.~\ref{fig:setup}.
Mathematically, this 1D projection is obtained by taking the line integral of the original 2D distribution $f(x,y)$ along the $Y$ direction at the position $X$, as
\begin{equation}
\label{Radon}
g_\theta(X) = \int_{-\infty}^\infty f(x,y)dY,
\end{equation}
which is widely known as the Radon transform\cite{Radon1986}.

In many wire scanners, the wire rotates around the $Y$~axis, so the scan is in the form of a cylindrical surface displaced from the $(x,y)$ plane.
However, in most cases, this displacement is negligible because the rotation radius is sufficiently larger than the beam size.
In each rotation, the wire crosses the beam twice, and a set of projections $g_\theta(X)$ is obtained for positions $z_1$ and $z_2$, respectively. 
Unlike the straight wire, a helical wire crosses the beam from different directions at $z_1$ and $z_2$, at right angles to each other (Fig.~\ref{fig:setup}(c))\cite{Seely2008}.
In this case, $g_\theta(X)$ and $g_\theta(Y)$ are simultaneously obtained, enabling the quasi $XY$ profile monitoring of the beam if the beam has a Gaussian shape and change in the beam profile while traveling from $z_1$ to $z_2$ is negligible.  
However, we demonstrate that this change is not negligible, but rather significant enough to determine the phase-space distribution of the beam.

\subsection{Experimental methods}
Figure~\ref{fig:setup}(b) shows a schematic illustration of the experimental setup.
We used a beam profile monitor (BPM-80, National Electrostatic Corp.) that has a helical wire continuously rotating at 19~Hz about the motor axis. 
In each wire rotation, it gives the $X$- and $Y$- projections of the beam in the form of the beam current intensity as a function of the position of the rotating wire.  
Figure~\ref{fig:setup}(d) shows a typical current signal from the wire.
The profile monitor is mounted on a rotary platform (Model ZRP100H, VacGen Ltd.) to observe 1D beam projections from $\theta=0$ to $2\pi$. 
This combined rotational motion enables the proposed dual-axis scanning of a wire, which is essential for the tomographic reconstruction of the 2D beam profile. 
Owing to the symmetry of the Radon transform, $g_\theta(X) = g_{\theta+\pi}(-X)$, and therefore, in principle, a 2D image can be reconstructed using projection data acquired only over the angular range from $0$ to $\pi$. However, in the present study, projections were measured over the full $0$ to $2\pi$ range to achieve more accurate and robust reconstruction.

The following sections present the results of beam experiments conducted using 7 and 5~keV Ar$^+$ from a duoplasmatron ion source. 
The beam is transported into the profile monitor through an electrostatic quadrupole triplet lens and a mask section. 
The Faraday cup at the end of the beamline measures the beam intensity. 
In the present setup as a test bench, the entire apparatus downstream of the rotary platform must be rotated, which takes on the order of 10 minutes to acquire all the data.
For practical use in beamline applications, a motorized dual-axis rotation mechanism is currently being developed, as discussed in a later section.

\begin{figure} [t]
\includegraphics[width=\linewidth]{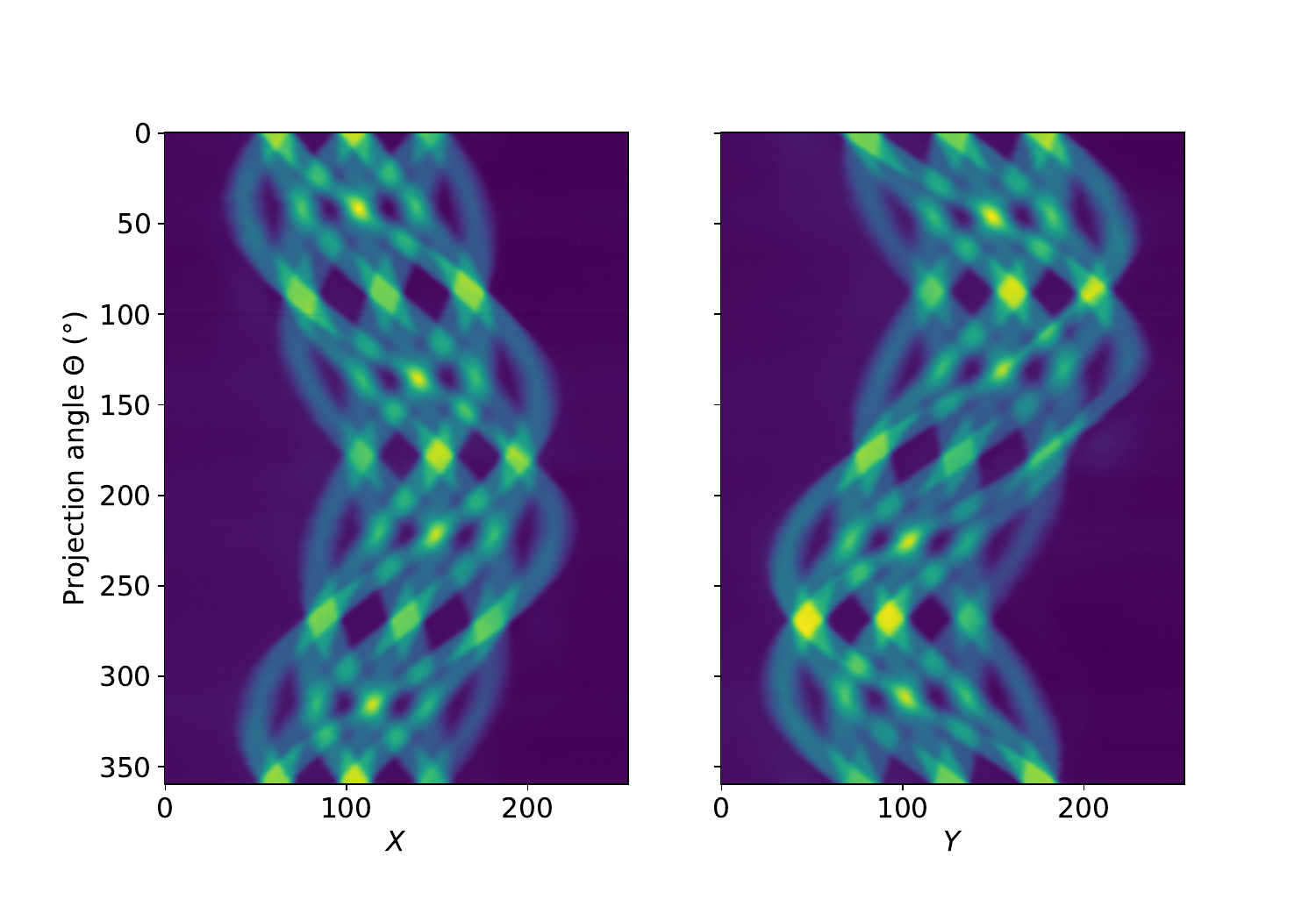}
\caption{
Observed data in the sinogram representation for $z_1$ (left) and $z_2$ (right). The horizontal axes represent the $X$ and $Y$ positions, respectively, while the vertical axis corresponds to the projection angle $\theta$. The baseline is subtracted from each 1D spectrum. The full horizontal range of 256 pixels corresponds to 37.3 mm.
}
\label{fig:sino}
\end{figure}

We performed several measurements to evaluate the image accuracy and position resolution of the present method, as well as the feasibility of evaluating the phase-space distribution of the beam using the two images $f_1(x,y)$ and $f_2(x,y)$. 
These measurements were done using a mask having a $3~\times~3$ grid of 2~mm square holes at 5~mm pitch.
The distance between the mask and the center of the profile monitor was approximately 245 mm. Figure~\ref{fig:sino} shows an example of a set of 1D projection data, referred to as a `sinogram', acquired over projection angles from $0$ to $2\pi$ in 1-degree steps.
We also used a mask with a \textit{fleur-de-lis} shape to demonstrate the reconstruction of a more complex 2D image. Finally, we measured the 2D profile of the ion beam without the mask under different beam transport conditions.

\subsection{Reconstruction algorithms}
In tomography, back projection is the most basic method for reconstructing a 2D image from multiple 1D projections. 

Back projetion, defined as $\int g_\theta(X) d\theta$, works by projecting each measured 1D profile $g_\theta(X)$ back over the 2D image space along its projection angle $\theta$, and summing all the contributions to reconstruct the 2D image $f(x,y)$. 

However, in general, this simple back projection produces blurred images, so more advanced algorithms are used to obtain clearer results.
Such reconstruction techniques can be broadly categorized into analytical and iterative methods.
Analytical methods use the properties of Fourier transforms, whereas iterative reconstruction repeatedly refines assumed projections to better match the measured projection data.
In this study, we performed image reconstruction using analytical methods including the Fourier transform (FT) and filtered back projection (FBP), as well as iterative methods such as maximum likelihood expectation maximization (ML-EM) and order subsets expectation maximization (OS-EM).
Analytical methods generally provide faster reconstruction times but are prone to the appearance of radial artifacts. ML-EM produces higher-accuracy reconstructed images with fewer artifacts but has a longer reconstruction time compared with analytical methods. Relative to ML-EM, OS-EM achieves shorter reconstruction times without significantly compromising the quality of the reconstructed images.
These methods are well established, and detailed descriptions can be found in the relevant literature\cite{A.C.Kak1999}. Below, we provide a brief overview of the methods employed and describe how each was applied in this study.

FT method is one of the most direct algorithms to reconstruct the original image $f(x,y)$ from the set of 1D projections $g_\theta(X)$. 
By collecting 1D Fourier transformation of the projections, namely, $\mathcal{F}(g_\theta(X))$ from all direction over $2\pi$, one can obtain the complete 2D transform $\mathcal{F}(f(x,y))$ according to the central slice theorem\cite{A.C.Kak1999}. 
Then, its 2D inverse Fourier transform recovers the original intensity distribution function $f(x,y)$.

FBP offers a more robust and computationally efficient reconstruction, and is therefore the most widely used algorithm in clinical CT imaging. 
It reconstructs the original distribution $f(x,y)$ by back-projecting the acquired 1D projections. Before back projection, high-pass frequency filters (\textit{e.g.}, Ramp and Shepp-logan filters) are applied to each 1D projection $g_\theta(X)$. These filters enhance high-frequency components to emphasize edges and fine structures in the reconstructed image.

The ML-EM algorithm is one of the standard iterative reconstruction methods, in which the reconstructed image is updated iteratively based on feedback from comparisons with the measured projection data\cite{Shepp1982}.
Figure~\ref{fig:ML-EM} illustrates the principle of image reconstruction using the ML-EM algorithm. 
Here, $k$ is the number of iterations, and $j$ and $i$ are the pixel indices of the 2D image data and the sinogram, respectively. 
First, prepare the initial image $x_j^{(0)}$ and calculate its projection $P_i^{(0)}$, which is then compared with the measured projection data $y_i$.  
The ratio $y_i/{P_i}^{(0)}$ is then back-projected to obtain a feedback factor $g_j^{(0)}$, which is multiplied to the initial image $x_j^{(0)}$ to generate the next image estimate $x_j^{(1)}$.  
This iterative process is repeated until convergence is achieved.

\begin{figure}[tb]
\centering
\begin{tikzpicture}
[
every node/.style={outer sep=0.12cm, inner sep=0},
arrow/.style={-{Stealth[length=0.25cm]}, thick},
block/.style={rectangle, draw, minimum height = 1cm, 
minimum width=4.1cm, thick, outer sep = 0},
block1/.style={rectangle, draw, minimum height = 1cm, 
minimum width=3.5cm, thick, outer sep = 0},
smallblock/.style={rectangle, draw, minimum height = 1cm, 
minimum width=1cm, thick, outer sep = 0},
sum/.style={thick, circle, draw, inner sep=0,
minimum size=12pt, outer sep=0},
point/.style={radius=2pt}
]
     \node [font=\small,smallblock] (Pi){${P_i}^{(k)}$};
     \node [font=\small,smallblock]at (-1.6,-3) (gi){${g_j}^{(k)}$};
     \node [sum, left=2of Pi] (c){$\times$};
     \node [sum, right=.6of Pi] (d){$/$};
     \node [font=\small,smallblock, right=.5 of d] (yi) {$y_i$};
     \draw[arrow] (yi) -- (d);
     \node [font=\small,smallblock, left=.5 of c] (xj) {$x_j^{(k)}$};
     \draw[arrow] (xj) -- (c);
    \draw [arrow] (d) -- +(0, -1)|-(gi) ;
    \draw [arrow] (gi) -- +(-1.1, 0)-|(c);
    \node[font=\small,above left=25pt,fill=white,below=1 of d](e){${y_i}/{P_i^{(k)}}$};
    \node[font=\small,above left=25pt,fill=white,below=1 of c](f) {$x_j^{(k+1)}=x_j^{(k)}\times g_j^{(k)}$};
    \node[font=\footnotesize,above right=35pt,fill=white,minimum height=0.5cm,above=-.25 of e]{Compare};
    \node[font=\footnotesize,above right=35pt,fill=white,minimum height=0.5cm,above=-.25 of f]{Feedback};
    \node[font=\footnotesize]at (-1.5,.5){Projection};
    \node[font=\footnotesize,above=.1of xj]{\shortstack{Estimated\\image}};
    \node[font=\footnotesize, above=.1of yi] {\shortstack{Measured\\projection}};
    \node[font=\footnotesize]at (0.1,-2.5){Back projection};
\draw[arrow] (Pi) -- (d)node[font=\small,above, midway]{} ;
\draw[arrow] (c) -- (Pi);
\end{tikzpicture}
\caption{Flowchart of Maximum Likelihood Expectation Maximization (ML-EM) method. An iterative algorithm for image reconstruction, where the estimated image $x_j(k)$ is forward projected to obtain the predicted projections $P_i(k)$. The ratio between the measured data $y_i$ and the predicted projections is then back-projected and used to update the image estimate.}
\label{fig:ML-EM}
\end{figure}

\begin{figure*}[t]
\centering
    \includegraphics[width=\linewidth]{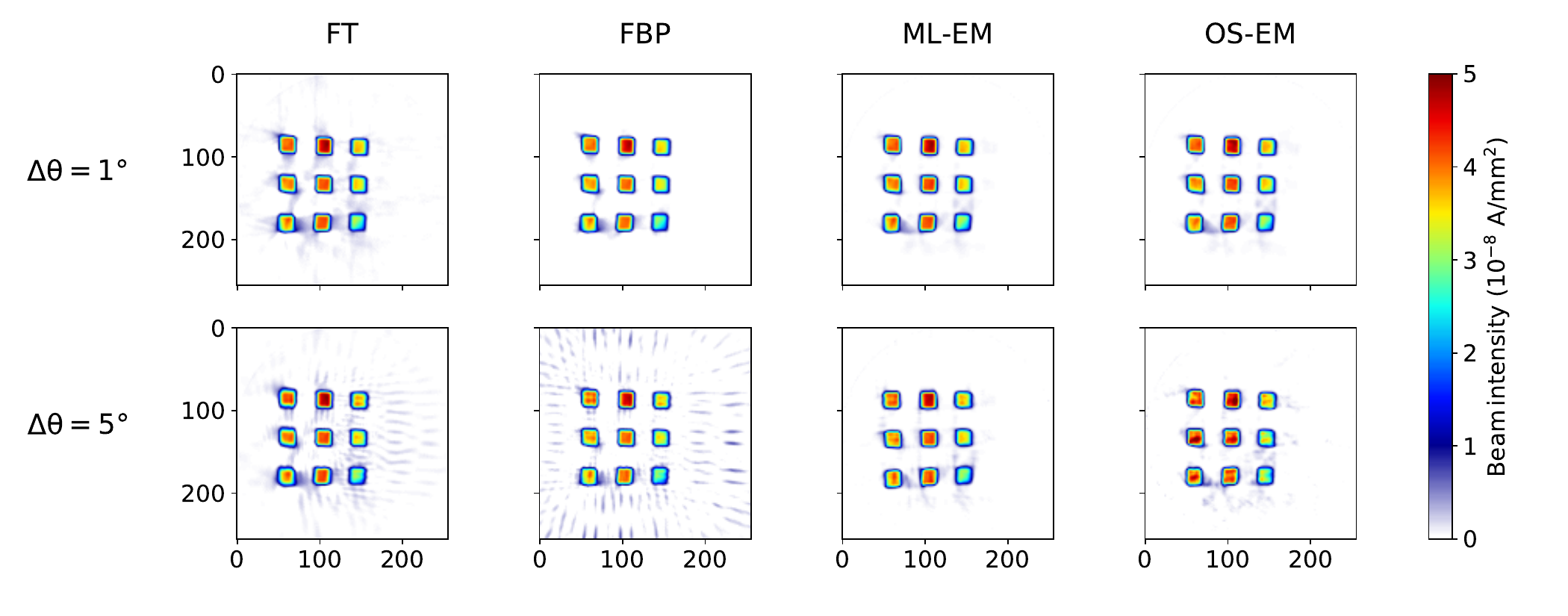}
    \caption{2D beam images reconstructed using different numbers of projections and reconstruction algorithms. 
    The upper images were reconstructed using 360 sets of projections measured at 1\textdegree~steps, and the lower ones were reconstructed using 72 sets at 5\textdegree~steps. From left to right, the columns show Fourier transform (FT), Filtered back projection (FBP), Maximum likelihood expectation maximization (ML-EM), and Ordered subset expectation maximization (OS-EM) results. The full range of 256 pixels corresponds to 37.3 mm. } 
    \label{fig:reconstructed image}
\end{figure*}

OS-EM is an accelerated variant of the ML-EM algorithm, in which the projection data are divided into ordered subsets, and the image is updated sequentially using each subset\cite{Hudson1994}. 
To ensure stable convergence and balanced image updates, it is generally recommended to divide the projection data into subsets of equal size and with uniformly distributed angles.
This approach significantly improves the convergence speed while preserving the statistical accuracy of the original ML-EM method.
The number of iterations required by ML-EM and OS-EM to achieve similar image quality is related by
\begin{align}
&(\text{Num. of ML-EM iterations}) \notag \\
&= (\text{Num. of subsets}) 
\times (\text{Num. of OS-EM iterations}).
\end{align}
Thus, increasing the number of subsets in OS-EM reduces the number of iterations.

\section{Result and discussions}
\label{experiment1}

\subsection{Comparison of different image reconstruction methods}
To obtain 1D projections of the beam intensity at various projection angles $\theta$ from 0 to $2\pi$, the platform's rotation angle was incremented in steps of $\Delta \theta = 1^\circ$, while continuously recording the 1D projection data.
After performing appropriate baseline subtraction, the acquired data were arranged into a 2D array as a sinogram (Fig.~\ref{fig:sino}).
Since the beam profile monitor measures the secondary electron yield from the wire, the electron emission coefficient must be taken into account.
We performed a calibration measurement to determine the effective electron emission coefficient per incident ion, defined as $\gamma_{\mathrm{eff}} = \int g_\theta(X)dX / I_{\mathrm{b}}$, where $I_{\mathrm{b}}$ is the beam current measured using the Faraday cup.
The 1D projection $g_\theta(X)$ was measured at various projection angles $\theta$ without using a mask.
The calibration yielded $\gamma_{\mathrm{eff}} = 0.22$ for a 5~keV Ar$^+$ beam, independent of the platform’s angle $\theta$.
This value is slightly smaller than the known electron emission coefficient of molybdenum\cite{Vance1967,Mahadevan1965,Medved1963,Ferron1981}, $\gamma \simeq 0.3$ , possibly due to incomplete electron collection efficiency in the beam profile monitor.
This calibration enables reconstruction of absolute 2D beam profiles in units of A/mm$^2$.

Using the set of 1D projection data obtained at position $z_1$, the 2D images of 7~keV Ar$^+$ were reconstructed using FT, FBP, ML-EM, and OS-EM algorithms, as shown in Fig.~\ref{fig:reconstructed image}. 
To enhance visual contrast and highlight structural features in the reconstructed images, the rainbow colormap was used throughout the paper.
The upper images were reconstructed using 360 projections measured at $\Delta \theta=1\degree$ steps; for the bottom ones, 72 projections at $\Delta \theta=5\degree$ steps.
The full range of 256 pixels corresponds to 37.3 mm in both $x$ and $y$.
In FT, the sinogram data were extended by adding a certain number of zeros to both ends of the 1D projections before FT. 
Called zero padding, this technique in FT helps improve image resolution and reduce artifacts.  
In FBP, we used a Ramp filter to reduce blurring and enhance image edges and fine structures. 
In ML-EM, the number of iterations was set to 60 to ensure sufficient convergence of the feedback factor.
In OS-EM, the number of subsets was 30 and the iteration count was 2, resulting in a total of 60 image updates.
    
All eight reconstructed images (Fig.~\ref{fig:reconstructed image}) reproduced the mask pattern well but exhibited different characteristics regarding accuracy and artifacts. 
Comparison of the 2D images reconstructed from 360 and 72 projections using all algorithms showed that using a larger number of projection angles produced clearer images.
In the case of the analytical methods, FT and FBP, radial artifacts that were prominent in the images reconstructed at 5\textdegree~steps largely decreased at 1\textdegree~steps.
In contrast, the iterative methods, ML-EM and OS-EM, showed fewer artifacts in both the 1\textdegree- and 5\textdegree-step images relative to the analytical methods.
Thus, iterative reconstruction is less prone to artifacts than analytical methods.
    
We also analyzed the sum of all pixels in the 256~$\times$~256 images, which should correspond to the beam current. 

The actual beam current during these measurements was determined to be 3.60~$\mu$A by integrating the 1D projection over the beam size, according to $\int g_\theta(X)dX$, using the calibration factor $\gamma_{\mathrm{eff}}$.
Table~\ref{Tab1} summarizes the beam currents obtained from integrating the reconstructed 2D images in Fig.~\ref{fig:reconstructed image}. 
These comparisons showed that the integrated intensities obtained from the FT and FBP methods varied with the angular step size $\Delta \theta$, and did not match the actual beam current.
In contrast, the integrated intensities from ML-EM and OS-EM were more accurate and independent of $\Delta \theta$.
This discrepancy reflects the fundamental difference between the two approaches: analytical methods apply the Fourier transform to the discretized projection data, while iterative methods refine the image through successive approximations to better match the measured data.

Regarding the image reconstruction time, the analytical methods, FT and FBP, were fast (on the order of several hundred milliseconds using a laptop computer with a 1.1~GHz Intel Core i5 processor); however, as mentioned earlier, the reconstructed images often had considerable artifacts. ML-EM had fewer artifacts in the reconstructed images but needed approximately 10~s for image reconstruction under the same environment. 
This was significantly longer than the time required by the analytical methods. 
OS-EM resulted in fewer artifacts and required only several hundred milliseconds for reconstruction, comparable to the reconstruction time of the analytical methods. Based on these results, it can be concluded that OS-EM is the most optimal reconstruction method for this measurement.

\begin{table}[h]
\centering
\caption{Integrated beam currents obtained from reconstructed images}
\begin{tabular}{lll}
\hline
Method\ \ \ \ \ \ \ \ \ & $\Delta\theta~=~$1\textdegree \ \ \ \ \ \ \ \ \ \ & $\Delta\theta~=~$5\textdegree\\
\hline
FT& $3.90~\rm{\mu A}$ &$4.00~\rm{\mu A}$\\
FBP & $3.30~\rm{\mu A}$ &$3.81~\rm{\mu A}$\\
ML-EM& $3.60~\rm{\mu A}$ & $3.60~\rm{\mu A}$ \\ 
OS-EM & $3.61~\rm{\mu A}$ & $3.68~\rm{\mu A}$\\
\hline
\end{tabular}
\label{Tab1}
\end{table}

\begin{figure}[tbp]
\centering
    \includegraphics[width=\linewidth]{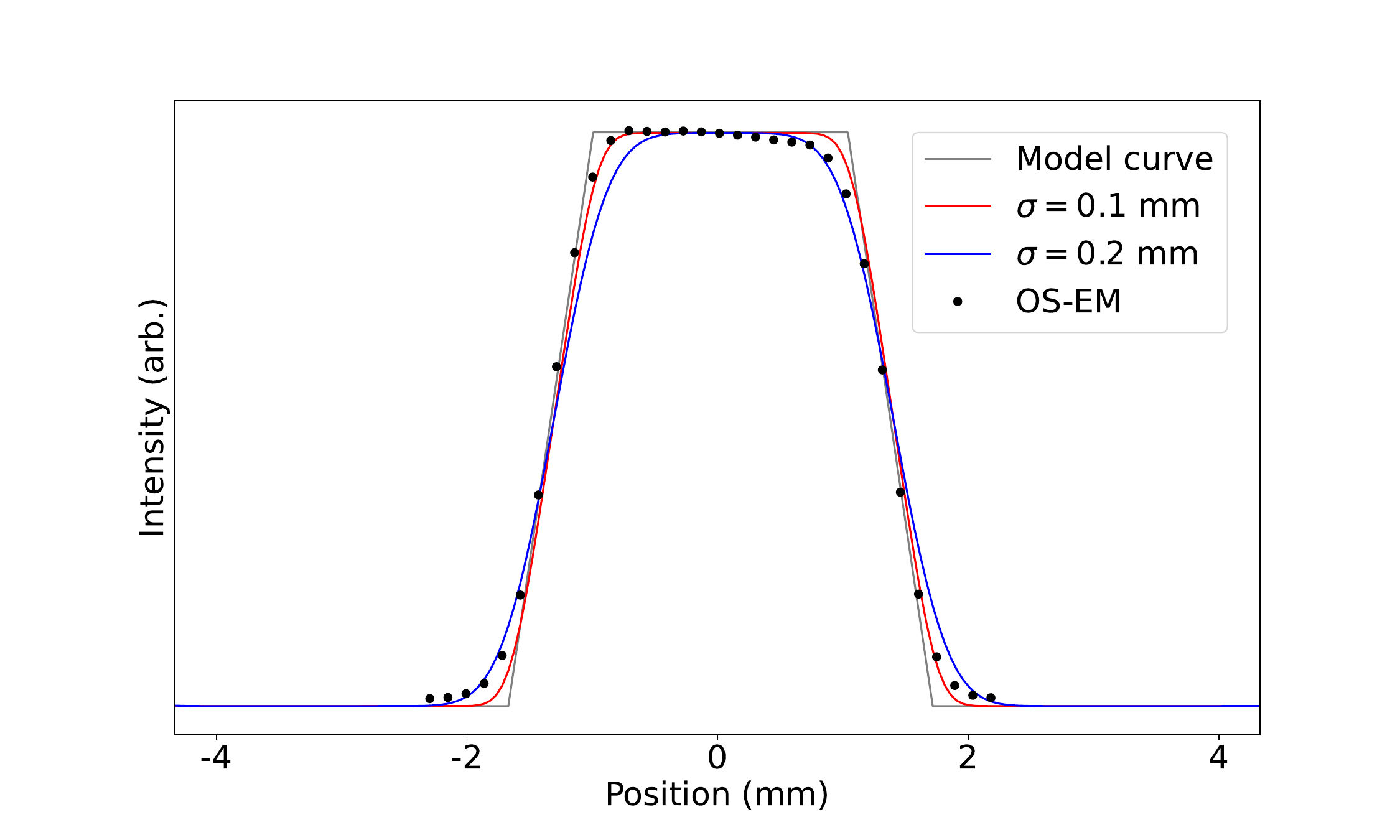}
    \caption{The $y$ projection of a single square from the 2D lattice pattern using OS-EM at $\Delta \theta = 1 \degree$. The solid gray line shows the model curve of the mask shape probed by a wire with a diameter of 0.5~mm. The red and blue lines are obtained by convolving the model curve with Gaussian functions with a standard deviation $\sigma=0.1$ and $0.2~\rm{mm}$, respectively.}
    \label{fig:dig}
\end{figure}

\begin{figure}[t]
\includegraphics[width=\linewidth]{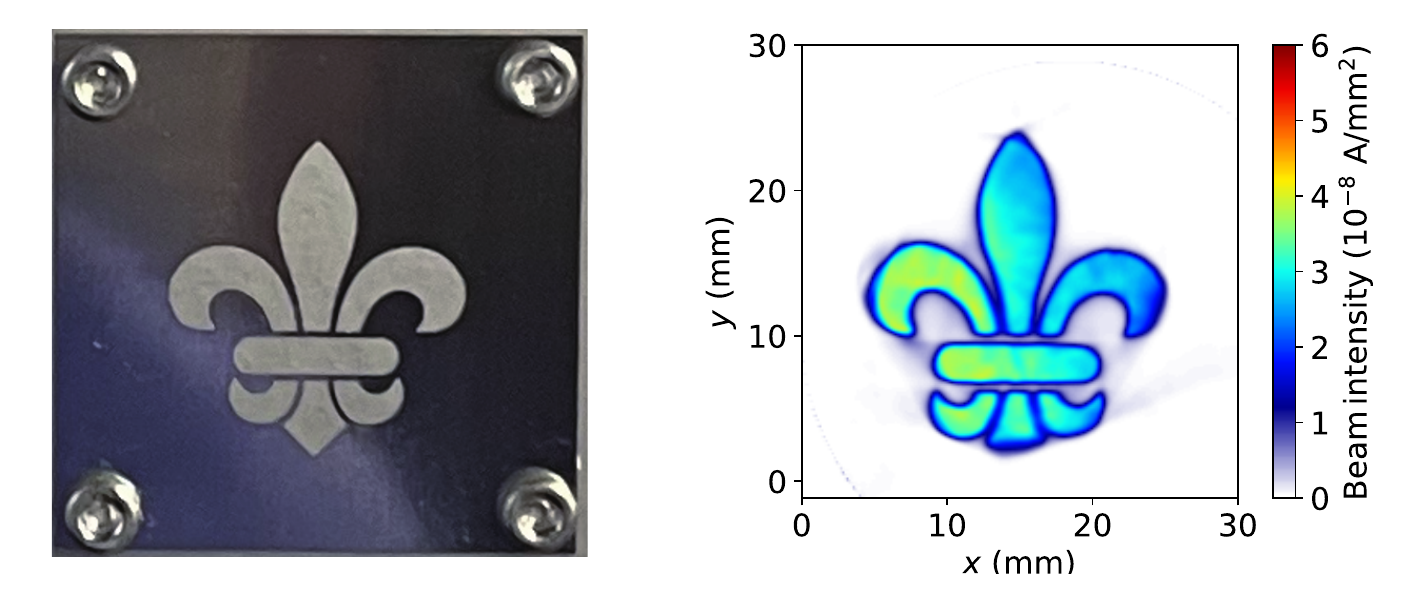}
\caption{Reconstruction of a 2D beam profile using the \textit{fleur-de-lis} mask. The original shape of the mask (left) and the 2D image of the beam reconstructed using OS-EM (right).}
\label{fig:rikkyo}
\end{figure}

\subsection{Spatial resolution}
The 1D position resolution of the beam profile monitor is inherently determined by the diameter of the wire, which is 0.5~mm in the present experiment. 
The actual spatial resolution may be limited by the sampling rate of the data acquisition system. In the present measurement, however, the sampling resolution was on the order of $1~\rm{\mu m/Sa}$, which is negligibly small compared to the wire diameter.
To evaluate the degradation of spatial resolution during the reconstruction process, the reconstructed shape of the masked beam was compared with a simulated profile. 
In the simulation, the rectangular shape of the mask was convolved with the wire diameter, resulting in a trapezoidal curve in the projection. 
This is shown by the gray curve in Fig.~\ref{fig:dig}, along with the 1D projection of the central spot of the $3~\times~3$ square pattern reconstructed using the OS-EM method ($\Delta \theta=1 \degree$).
The red and blue curves represent the simulation obtained by further convolving the model curve with Gaussian function of standard deviations $\sigma=0.1$ and $0.2~\rm{mm}$, respectively, to account for the degradation of position resolution.
From this comparison, the degree of Gaussian blurring in the present measurement is estimated to lie between 0.1 and 0.2~mm.
Since the current analysis does not take into account the image blurring caused by beam divergence between the mask and the wire, the estimated resolution degradation of 0.1 to 0.2~mm should be regarded as an upper limit in the present tomographic reconstruction method.
Depending on the characteristic of the beam shape, the blurring may be reduced by selecting the appropriate reconstruction algorithm and related parameters, such as the number of iterations. 

We also examined the reconstruction of the 2D beam profile using the mask having a \textit{fleur-de-lis} shape instead of the $3~\times~3$ grid. 
Figure~\ref{fig:rikkyo} shows the original shape of the mask (left) and the reconstructed 2D image of the beam (right).
Reconstruction was performed via OS-EM from 360~projections with the same number of subsets and iterations as those used for the grid pattern. 
The \textit{fleur-de-lis} pattern, despite its intricate design, was reconstructed with remarkable accuracy. Even features narrower than the wire diameter, such as the 0.43~mm gaps, were distinctly resolved.
The nonuniformity and the cutoff at the bottom of the image were not caused by the reconstruction process; they were inherent to the original beam shape.
Thus, the proposed method is applicable to various beams with complex and fine structures, such as spike-like features, and high-resolution 2D imaging using particle beams.

\begin{figure}[tb]
\begin{minipage}{0.49\hsize}
\centering
\includegraphics[scale=.22]{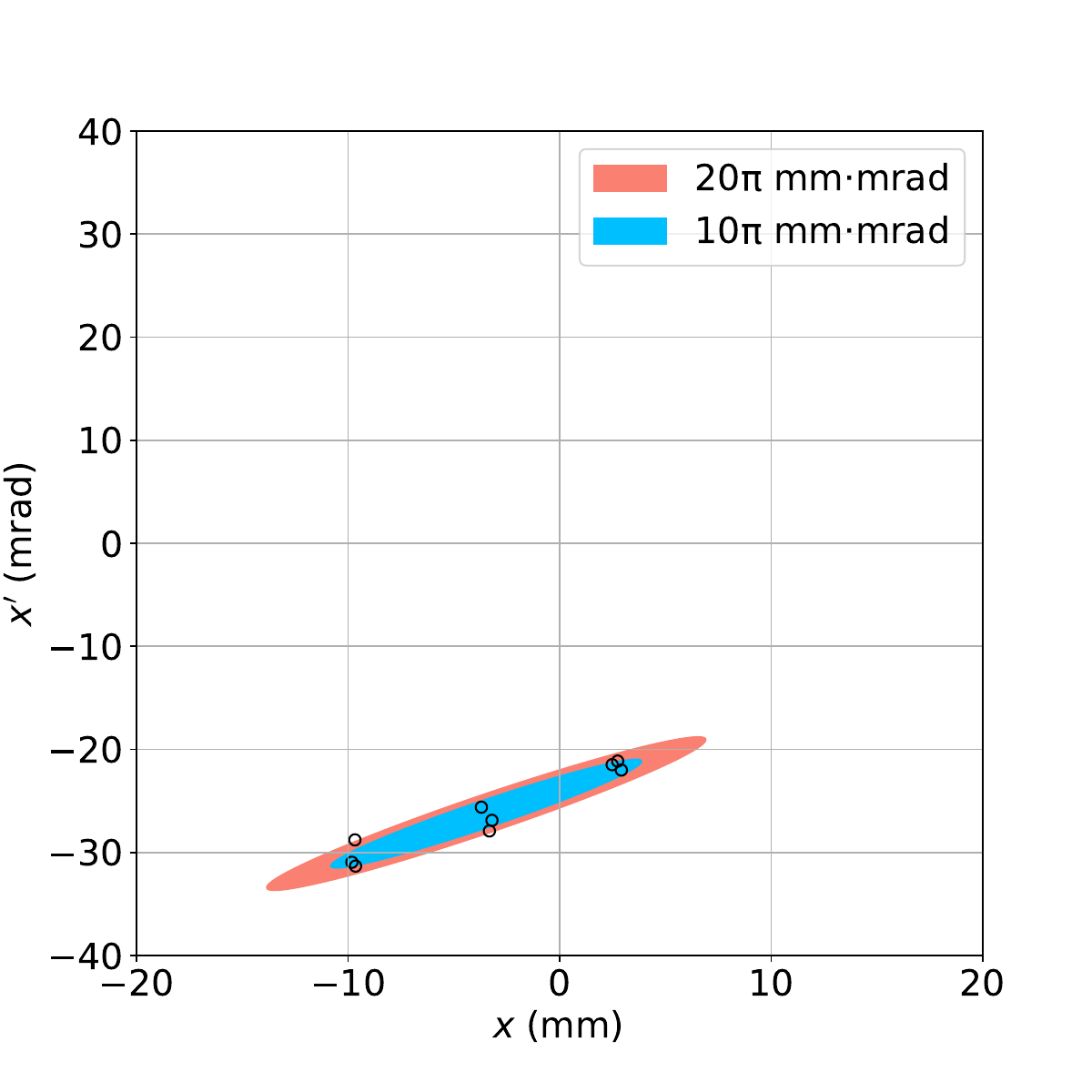}
\end{minipage}
\begin{minipage}{0.49\hsize}
\centering
\includegraphics[scale=.22]{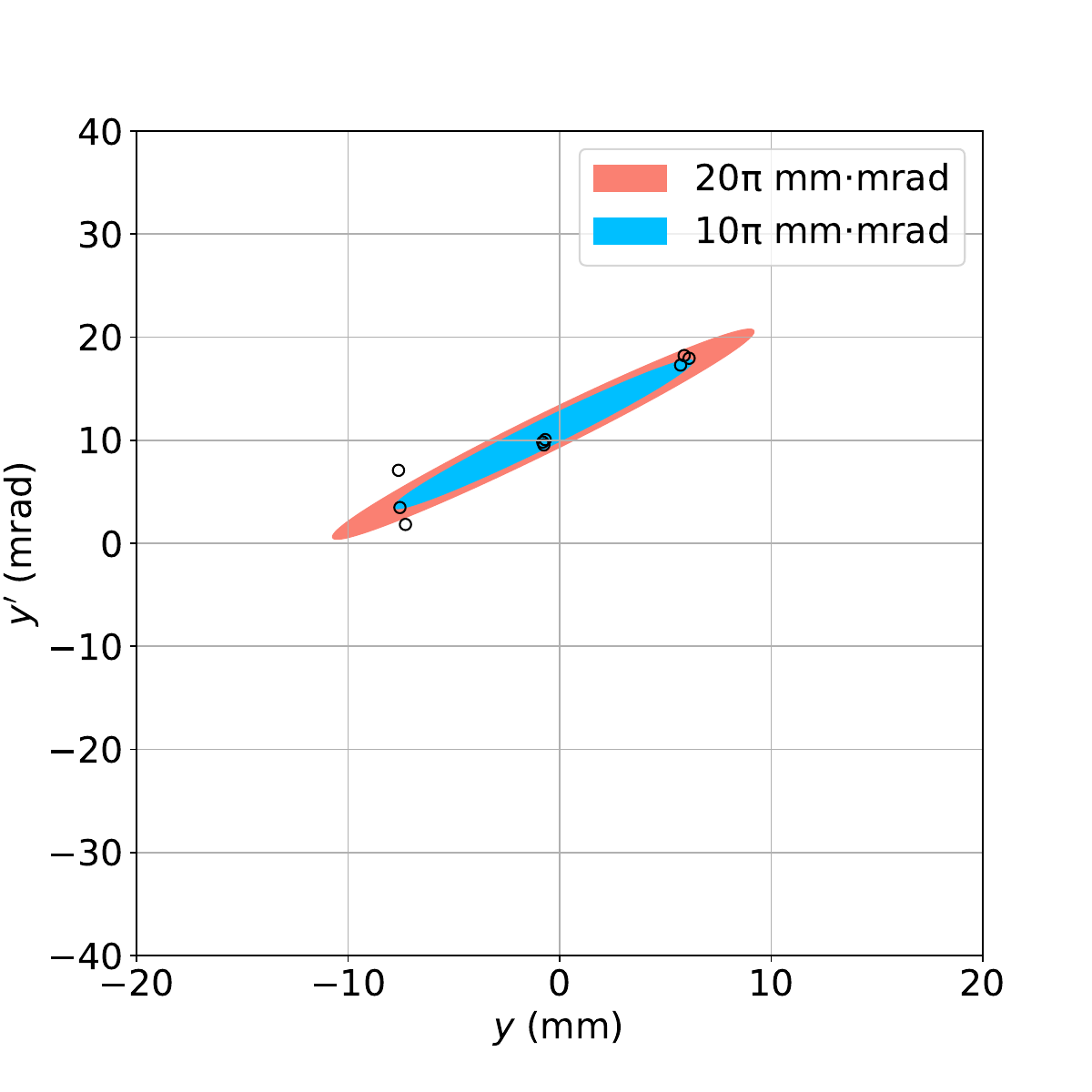}
\end{minipage}
\caption{Phase-space distribution of the beam obtained from the reconstructed 2D images in the $3~\times~3$ grid pattern at $z_1$ and $z_2$. The filled area represents an ellipses assuming a beam emittances of $10\pi$ and $20\pi$~mm$\cdot$mrad, respectively. }
\label{fig:psd}
\end{figure}

\subsection{Beam axis and emittance measurement}
As depicted in Fig.~\ref{fig:setup}(c), the wire crosses the beam once at $z=z_1$ and then again at $z_2$.  
In the present experiment, $z_1$ and $z_2$ were separated by 54~mm along the $z$~axis. 
This feature can be used to obtain the 2D beam profiles simultaneously at $z_1$ and $z_2$, which can provide the phase-space information of the beam.  
From the centroid positions of each spot in the $3~\times~3$ grid pattern, the relationship between the positions and the beam angle from the $z$~axis was obtained and plotted in Fig.~\ref{fig:psd} left and right for $x$ and $y$, respectively. 

From this plot, it is evident that the beam is off-aligned  from the $z$~axis approximately -25 and +10~mrad in the $x$ and $y$ directions, respectively. 
The beam’s phase-space distribution exhibited an upward-right tilt, indicating that the beam is diverging rather than converging. The angular spreads were approximately 10 and 20~mrad in the $x$ and $y$ axes, respectively. 
The beam emittance could be obtained from the phase-space distribution by analyzing the area occupied by the beam.
Although the number of plotted points is not enough to estimate the accurate emittance of the beam, reference ellipses with areas of $10\pi$ and $20\pi$~mm$\cdot$mrad are overlaid on the data, which reasonably explained the measured phase-space distribution of the beam. 
The use of masks with finer structures, such as the pepper pot technique \cite{Kremers2013}, would give a more detailed phase-space profile of the beam.  

\begin{figure}[tb]
\includegraphics[width=\linewidth]{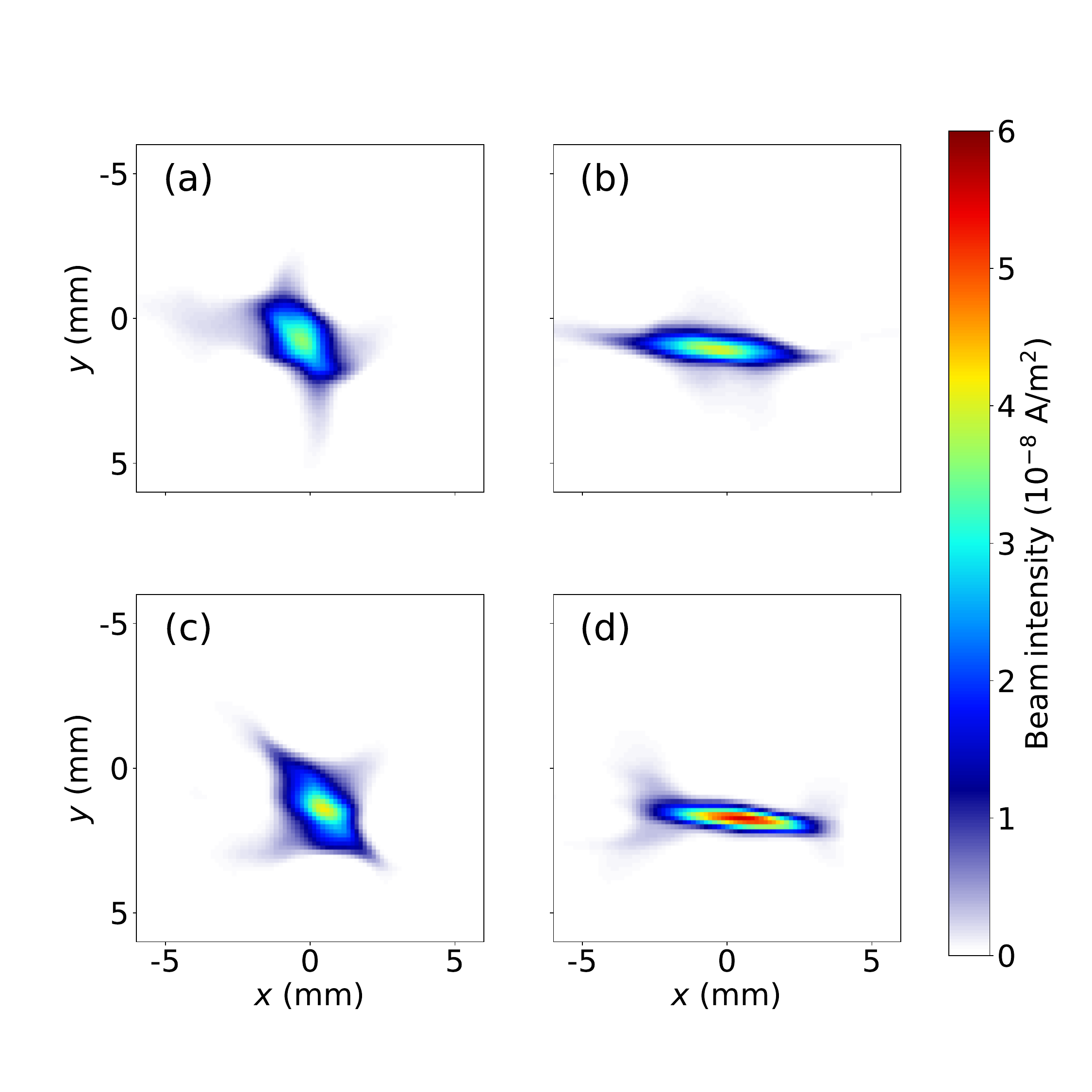}

\caption{2D profiles of (a) a point-focused beam and (b) a flat beam measured at $z=z_1$. The reconstruction method was OS-EM. The images for the same beams observed at $z_2$ are also shown in (c) and (d), respectively. }

\label{reconstructed img nomask}
\end{figure}

\subsection{Measurements of the focused-beam profile} \label{Experiment2}
To demonstrate the practical applicability, 2D beam profile measurements were conducted without shaping the beam using a mask. The measurements were performed using a 5~keV~$\rm{Ar^+}$ beam transported through an electrostatic quadrupole triplet lens. Before entering the lens, the beam passed through a slit having 4~mm and 5~mm opening in $x$ and $y$, respectively.
Figures~\ref{reconstructed img nomask}(a) and \ref{reconstructed img nomask}(b) show 2D beam profiles reconstructed using OS-EM from the 1D spectrum obtained at $z_1$.
The numbers of subsets and iterations were 30 and 2, respectively. 
In Fig.~\ref{reconstructed img nomask}(a), the lens voltage was set to focus the beam on both the $x$ and $y$ axes as much as possible. 
In Fig.~\ref{reconstructed img nomask}(b), the lens voltage was set such that the beam diverged in the $x$-axis direction while converging in the $y$-axis direction.
The characteristics of the beam shapes under the two lens settings were clear in the reconstructed 2D images.
Figures~\ref{reconstructed img nomask}(c) and \ref{reconstructed img nomask}(d) show the corresponding 2D beam profiles observed at $z_2$.

The beam directions were determined from the shift of the intensity centroids between the images at $z_1$ and $z_2$. 
For the point-focused beam, the deviation of the beam direction from the z-axis was $11.2~\rm{mrad}$ in the $x$ direction and $11.6~\rm{mrad}$ in the $y$ direction. 
As demonstrated, a notable advantage of this method is its ability to provide not only a high-resolution 2D beam profile, but also the beam direction simultaneously with milliradian precision.

\section{Summary}
In this work, we proposed the use of a novel dual-axis rotating wire to observe the 1D projection of a beam from $2\pi$ direction surrounding the beam. 
Through computed tomography, the 2D profile of a beam was successfully reconstructed from an observed sinogram using different algorithms: FT, FBP, ML-EM, and OS-EM. 
From the perspectives of accuracy, artifact reduction, and processing time, OS-EM was found to be the most suitable for the present case.
The resolution degradation caused by reconstruction was evaluated by analyzing the mask pattern, which was found to be less than 0.2~mm in the present case.
Furthermore, in this demonstration, the 2D profile of a beam obtained through a complex \textit{fleur-de-lis}-shaped mask was reconstructed with high accuracy. 
In addition to real-space 2D imaging, the proposed method enables measurement of the phase-space distribution of the beam using the 2D beam profile at two positions, $z_1$ and $z_2$, along the beam axis. 
To demonstrate its practical use as a beam profiler, we investigated changes in the 2D beam shape by varying the parameters of the quadrupole lenses in the beamline.
It is worth emphasizing that all abovementioned 2D profiles were obtained in absolute intensity units (A/mm$^2$). 
Most of the necessary beam diagnosis in usual beam experiments and applications, namely, absolute 2D profiling and phase-space distribution estimation, are achieved using a single robust device. 
Given its wide dynamic range, from the subnanoampere level to the kilowatt class, the proposed method is expected to have a broad range of applications.
This novel dual-axis rotating-wire mechanism can be achieved using a simple bevel gear system, eliminating the need to rotate the entire beamline (as in the present experiment).
By integrating fully motorized equipment, the data acquisition time is expected to be reduced from approximately 10 minutes to just a few seconds. 
With such a setup, rapid 2D image reconstruction will enable movie-like visualization of the beam profile. To achieve this, FPGA-based pipelined signal processing is well suited for real-time data handling.

\section*{Data Availability Statement}
The data that support the findings of this study are available from the corresponding author upon reasonable request.

\section*{Acknowledgment}
This work was, in part, supported by the JSPS KAKENHI (Grant Nos. 20H05844, 20H05848, and 25H00610).

\section*{References}
\bibliography{BPM}

\begin{thebibliography}{18}%
\makeatletter
\providecommand \@ifxundefined [1]{%
 \@ifx{#1\undefined}
}%
\providecommand \@ifnum [1]{%
 \ifnum #1\expandafter \@firstoftwo
 \else \expandafter \@secondoftwo
 \fi
}%
\providecommand \@ifx [1]{%
 \ifx #1\expandafter \@firstoftwo
 \else \expandafter \@secondoftwo
 \fi
}%
\providecommand \natexlab [1]{#1}%
\providecommand \enquote  [1]{``#1''}%
\providecommand \bibnamefont  [1]{#1}%
\providecommand \bibfnamefont [1]{#1}%
\providecommand \citenamefont [1]{#1}%
\providecommand \href@noop [0]{\@secondoftwo}%
\providecommand \href [0]{\begingroup \@sanitize@url \@href}%
\providecommand \@href[1]{\@@startlink{#1}\@@href}%
\providecommand \@@href[1]{\endgroup#1\@@endlink}%
\providecommand \@sanitize@url [0]{\catcode `\\12\catcode `\$12\catcode `\&12\catcode `\#12\catcode `\^12\catcode `\_12\catcode `\%12\relax}%
\providecommand \@@startlink[1]{}%
\providecommand \@@endlink[0]{}%
\providecommand \url  [0]{\begingroup\@sanitize@url \@url }%
\providecommand \@url [1]{\endgroup\@href {#1}{\urlprefix }}%
\providecommand \urlprefix  [0]{URL }%
\providecommand \Eprint [0]{\href }%
\providecommand \doibase [0]{http://dx.doi.org/}%
\providecommand \selectlanguage [0]{\@gobble}%
\providecommand \bibinfo  [0]{\@secondoftwo}%
\providecommand \bibfield  [0]{\@secondoftwo}%
\providecommand \translation [1]{[#1]}%
\providecommand \BibitemOpen [0]{}%
\providecommand \bibitemStop [0]{}%
\providecommand \bibitemNoStop [0]{.\EOS\space}%
\providecommand \EOS [0]{\spacefactor3000\relax}%
\providecommand \BibitemShut  [1]{\csname bibitem#1\endcsname}%
\let\auto@bib@innerbib\@empty
\bibitem [{\citenamefont {Seely}\ \emph {et~al.}(2008)\citenamefont {Seely}, \citenamefont {Bruhns}, \citenamefont {Savin}, \citenamefont {Kvale}, \citenamefont {Galutschek}, \citenamefont {Aliabadi},\ and\ \citenamefont {Havener}}]{Seely2008}%
  \BibitemOpen
  \bibfield  {author} {\bibinfo {author} {\bibfnamefont {D.}~\bibnamefont {Seely}}, \bibinfo {author} {\bibfnamefont {H.}~\bibnamefont {Bruhns}}, \bibinfo {author} {\bibfnamefont {D.}~\bibnamefont {Savin}}, \bibinfo {author} {\bibfnamefont {T.}~\bibnamefont {Kvale}}, \bibinfo {author} {\bibfnamefont {E.}~\bibnamefont {Galutschek}}, \bibinfo {author} {\bibfnamefont {H.}~\bibnamefont {Aliabadi}}, \ and\ \bibinfo {author} {\bibfnamefont {C.}~\bibnamefont {Havener}},\ }\href {\doibase 10.1016/j.nima.2007.10.041} {\bibfield  {journal} {\bibinfo  {journal} {Nucl. Instruments Methods Phys. Res. Sect. A Accel. Spectrometers, Detect. Assoc. Equip.}\ }\textbf {\bibinfo {volume} {585}},\ \bibinfo {pages} {69} (\bibinfo {year} {2008})}\BibitemShut {NoStop}%
\bibitem [{\citenamefont {Harryman}\ and\ \citenamefont {Wilcox}(2017)}]{Harryman2017}%
  \BibitemOpen
  \bibfield  {author} {\bibinfo {author} {\bibfnamefont {D.~M.}\ \bibnamefont {Harryman}}\ and\ \bibinfo {author} {\bibfnamefont {C.~C.}\ \bibnamefont {Wilcox}},\ }\href {\doibase 10.18429/JACoW-IBIC2017-WEPCC18} {\bibfield  {journal} {\bibinfo  {journal} {Proc. 6th Int. Beam Instrum. Conf. IBIC 2017}\ ,\ \bibinfo {pages} {396}} (\bibinfo {year} {2017})}\BibitemShut {NoStop}%
\bibitem [{\citenamefont {Arutunian}\ \emph {et~al.}(2021)\citenamefont {Arutunian}, \citenamefont {Margaryan}, \citenamefont {Harutyunyan}, \citenamefont {Lazareva}, \citenamefont {Darpasyan}, \citenamefont {Gyulamiryan}, \citenamefont {Chung},\ and\ \citenamefont {Kwak}}]{Arutunian2021a}%
  \BibitemOpen
  \bibfield  {author} {\bibinfo {author} {\bibfnamefont {S.~G.}\ \bibnamefont {Arutunian}}, \bibinfo {author} {\bibfnamefont {A.~V.}\ \bibnamefont {Margaryan}}, \bibinfo {author} {\bibfnamefont {G.~S.}\ \bibnamefont {Harutyunyan}}, \bibinfo {author} {\bibfnamefont {E.~G.}\ \bibnamefont {Lazareva}}, \bibinfo {author} {\bibfnamefont {A.~T.}\ \bibnamefont {Darpasyan}}, \bibinfo {author} {\bibfnamefont {D.~S.}\ \bibnamefont {Gyulamiryan}}, \bibinfo {author} {\bibfnamefont {M.}~\bibnamefont {Chung}}, \ and\ \bibinfo {author} {\bibfnamefont {D.}~\bibnamefont {Kwak}},\ }\href {\doibase 10.1063/5.0028666} {\bibfield  {journal} {\bibinfo  {journal} {Rev. Sci. Instrum.}\ }\textbf {\bibinfo {volume} {92}} (\bibinfo {year} {2021}),\ 10.1063/5.0028666}\BibitemShut {NoStop}%
\bibitem [{\citenamefont {Igarashi}\ \emph {et~al.}(2002)\citenamefont {Igarashi}, \citenamefont {Arakawa}, \citenamefont {Koba}, \citenamefont {Sato}, \citenamefont {Toyama},\ and\ \citenamefont {Yoshii}}]{Igarashi2002}%
  \BibitemOpen
  \bibfield  {author} {\bibinfo {author} {\bibfnamefont {S.}~\bibnamefont {Igarashi}}, \bibinfo {author} {\bibfnamefont {D.}~\bibnamefont {Arakawa}}, \bibinfo {author} {\bibfnamefont {K.}~\bibnamefont {Koba}}, \bibinfo {author} {\bibfnamefont {H.}~\bibnamefont {Sato}}, \bibinfo {author} {\bibfnamefont {T.}~\bibnamefont {Toyama}}, \ and\ \bibinfo {author} {\bibfnamefont {M.}~\bibnamefont {Yoshii}},\ }\href {\doibase 10.1016/s0168-9002(01)01515-7} {\bibfield  {journal} {\bibinfo  {journal} {Nucl. Instruments Methods Phys. Res. Sect. A Accel. Spectrometers, Detect. Assoc. Equip.}\ }\textbf {\bibinfo {volume} {482}},\ \bibinfo {pages} {32} (\bibinfo {year} {2002})}\BibitemShut {NoStop}%
\bibitem [{\citenamefont {Stratakis}\ \emph {et~al.}(2006)\citenamefont {Stratakis}, \citenamefont {Kishek}, \citenamefont {Li}, \citenamefont {Bernal}, \citenamefont {Walter}, \citenamefont {Quinn}, \citenamefont {Reiser},\ and\ \citenamefont {O'Shea}}]{Stratakis2006}%
  \BibitemOpen
  \bibfield  {author} {\bibinfo {author} {\bibfnamefont {D.}~\bibnamefont {Stratakis}}, \bibinfo {author} {\bibfnamefont {R.~A.}\ \bibnamefont {Kishek}}, \bibinfo {author} {\bibfnamefont {H.}~\bibnamefont {Li}}, \bibinfo {author} {\bibfnamefont {S.}~\bibnamefont {Bernal}}, \bibinfo {author} {\bibfnamefont {M.}~\bibnamefont {Walter}}, \bibinfo {author} {\bibfnamefont {B.}~\bibnamefont {Quinn}}, \bibinfo {author} {\bibfnamefont {M.}~\bibnamefont {Reiser}}, \ and\ \bibinfo {author} {\bibfnamefont {P.~G.}\ \bibnamefont {O'Shea}},\ }\href {\doibase 10.1103/PhysRevSTAB.9.112801} {\bibfield  {journal} {\bibinfo  {journal} {Phys. Rev. Spec. Top. - Accel. Beams}\ }\textbf {\bibinfo {volume} {9}},\ \bibinfo {pages} {1} (\bibinfo {year} {2006})}\BibitemShut {NoStop}%
\bibitem [{\citenamefont {McKee}, \citenamefont {O'Shea},\ and\ \citenamefont {Madey}(1995)}]{McKee1995}%
  \BibitemOpen
  \bibfield  {author} {\bibinfo {author} {\bibfnamefont {C.}~\bibnamefont {McKee}}, \bibinfo {author} {\bibfnamefont {P.}~\bibnamefont {O'Shea}}, \ and\ \bibinfo {author} {\bibfnamefont {J.}~\bibnamefont {Madey}},\ }\href {\doibase 10.1016/0168-9002(94)01411-6} {\bibfield  {journal} {\bibinfo  {journal} {Nucl. Instruments Methods Phys. Res. Sect. A Accel. Spectrometers, Detect. Assoc. Equip.}\ }\textbf {\bibinfo {volume} {358}},\ \bibinfo {pages} {264} (\bibinfo {year} {1995})}\BibitemShut {NoStop}%
\bibitem [{\citenamefont {Xiang}\ \emph {et~al.}(2009)\citenamefont {Xiang}, \citenamefont {Du}, \citenamefont {Yan}, \citenamefont {Li}, \citenamefont {Huang}, \citenamefont {Tang},\ and\ \citenamefont {Lin}}]{Xiang2009}%
  \BibitemOpen
  \bibfield  {author} {\bibinfo {author} {\bibfnamefont {D.}~\bibnamefont {Xiang}}, \bibinfo {author} {\bibfnamefont {Y.-C.}\ \bibnamefont {Du}}, \bibinfo {author} {\bibfnamefont {L.-X.}\ \bibnamefont {Yan}}, \bibinfo {author} {\bibfnamefont {R.-K.}\ \bibnamefont {Li}}, \bibinfo {author} {\bibfnamefont {W.-H.}\ \bibnamefont {Huang}}, \bibinfo {author} {\bibfnamefont {C.-X.}\ \bibnamefont {Tang}}, \ and\ \bibinfo {author} {\bibfnamefont {Y.-Z.}\ \bibnamefont {Lin}},\ }\href {\doibase 10.1103/PhysRevSTAB.12.022801} {\bibfield  {journal} {\bibinfo  {journal} {Phys. Rev. Spec. Top. - Accel. Beams}\ }\textbf {\bibinfo {volume} {12}},\ \bibinfo {pages} {022801} (\bibinfo {year} {2009})}\BibitemShut {NoStop}%
\bibitem [{\citenamefont {Xing}\ \emph {et~al.}(2018)\citenamefont {Xing}, \citenamefont {Du}, \citenamefont {Guan}, \citenamefont {Tang}, \citenamefont {Wang}, \citenamefont {Wang},\ and\ \citenamefont {Zheng}}]{Xing2018}%
  \BibitemOpen
  \bibfield  {author} {\bibinfo {author} {\bibfnamefont {Q.~Z.}\ \bibnamefont {Xing}}, \bibinfo {author} {\bibfnamefont {L.}~\bibnamefont {Du}}, \bibinfo {author} {\bibfnamefont {X.~L.}\ \bibnamefont {Guan}}, \bibinfo {author} {\bibfnamefont {C.~X.}\ \bibnamefont {Tang}}, \bibinfo {author} {\bibfnamefont {M.~W.}\ \bibnamefont {Wang}}, \bibinfo {author} {\bibfnamefont {X.~W.}\ \bibnamefont {Wang}}, \ and\ \bibinfo {author} {\bibfnamefont {S.~X.}\ \bibnamefont {Zheng}},\ }\href {\doibase 10.1103/PhysRevAccelBeams.21.072801} {\bibfield  {journal} {\bibinfo  {journal} {Phys. Rev. Accel. Beams}\ }\textbf {\bibinfo {volume} {21}},\ \bibinfo {pages} {072801} (\bibinfo {year} {2018})}\BibitemShut {NoStop}%
\bibitem [{\citenamefont {Moretto-Capelle}\ \emph {et~al.}(2023)\citenamefont {Moretto-Capelle}, \citenamefont {Panader}, \citenamefont {Polizzi},\ and\ \citenamefont {Champeaux}}]{Moretto-Capelle2023}%
  \BibitemOpen
  \bibfield  {author} {\bibinfo {author} {\bibfnamefont {P.}~\bibnamefont {Moretto-Capelle}}, \bibinfo {author} {\bibfnamefont {E.}~\bibnamefont {Panader}}, \bibinfo {author} {\bibfnamefont {L.}~\bibnamefont {Polizzi}}, \ and\ \bibinfo {author} {\bibfnamefont {J.~P.}\ \bibnamefont {Champeaux}},\ }\href {\doibase 10.1063/5.0158663} {\bibfield  {journal} {\bibinfo  {journal} {Rev. Sci. Instrum.}\ }\textbf {\bibinfo {volume} {94}},\ \bibinfo {pages} {083306} (\bibinfo {year} {2023})}\BibitemShut {NoStop}%
\bibitem [{\citenamefont {Radon}(1986)}]{Radon1986}%
  \BibitemOpen
  \bibfield  {author} {\bibinfo {author} {\bibfnamefont {J.}~\bibnamefont {Radon}},\ }\href {\doibase 10.1109/tmi.1986.4307775} {\bibfield  {journal} {\bibinfo  {journal} {IEEE Trans. Med. Imaging}\ }\textbf {\bibinfo {volume} {MI-5}},\ \bibinfo {pages} {170} (\bibinfo {year} {1986})}\BibitemShut {NoStop}%
\bibitem [{\citenamefont {{a. C. Kak}}\ and\ \citenamefont {Stanley}(1999)}]{A.C.Kak1999}%
  \BibitemOpen
  \bibfield  {author} {\bibinfo {author} {\bibnamefont {{a. C. Kak}}}\ and\ \bibinfo {author} {\bibfnamefont {M.}~\bibnamefont {Stanley}},\ }\href@noop {} {\bibfield  {journal} {\bibinfo  {journal} {Princ. Comput. Tomogr. Imaging}\ ,\ \bibinfo {pages} {49}} (\bibinfo {year} {1999})}\BibitemShut {NoStop}%
\bibitem [{\citenamefont {Shepp}\ and\ \citenamefont {Vardi}(1982)}]{Shepp1982}%
  \BibitemOpen
  \bibfield  {author} {\bibinfo {author} {\bibfnamefont {L.~A.}\ \bibnamefont {Shepp}}\ and\ \bibinfo {author} {\bibfnamefont {Y.}~\bibnamefont {Vardi}},\ }\href {\doibase 10.1109/TMI.1982.4307558} {\bibfield  {journal} {\bibinfo  {journal} {IEEE Trans. Med. Imaging}\ }\textbf {\bibinfo {volume} {1}},\ \bibinfo {pages} {113} (\bibinfo {year} {1982})}\BibitemShut {NoStop}%
\bibitem [{\citenamefont {Hudson}\ and\ \citenamefont {Larkin}(1994)}]{Hudson1994}%
  \BibitemOpen
  \bibfield  {author} {\bibinfo {author} {\bibfnamefont {H.~M.}\ \bibnamefont {Hudson}}\ and\ \bibinfo {author} {\bibfnamefont {R.~S.}\ \bibnamefont {Larkin}},\ }\href@noop {} {\bibfield  {journal} {\bibinfo  {journal} {IEEE Trans. Med. Imaging}\ }\textbf {\bibinfo {volume} {13}},\ \bibinfo {pages} {601} (\bibinfo {year} {1994})}\BibitemShut {NoStop}%
\bibitem [{\citenamefont {Vance}(1967)}]{Vance1967}%
  \BibitemOpen
  \bibfield  {author} {\bibinfo {author} {\bibfnamefont {D.~W.}\ \bibnamefont {Vance}},\ }\href {\doibase 10.1103/PhysRev.164.372} {\bibfield  {journal} {\bibinfo  {journal} {Phys. Rev.}\ }\textbf {\bibinfo {volume} {164}},\ \bibinfo {pages} {372} (\bibinfo {year} {1967})}\BibitemShut {NoStop}%
\bibitem [{\citenamefont {Mahadevan}\ \emph {et~al.}(1965)\citenamefont {Mahadevan}, \citenamefont {Magnuson}, \citenamefont {Layton},\ and\ \citenamefont {Carlston}}]{Mahadevan1965}%
  \BibitemOpen
  \bibfield  {author} {\bibinfo {author} {\bibfnamefont {P.}~\bibnamefont {Mahadevan}}, \bibinfo {author} {\bibfnamefont {G.~D.}\ \bibnamefont {Magnuson}}, \bibinfo {author} {\bibfnamefont {J.~K.}\ \bibnamefont {Layton}}, \ and\ \bibinfo {author} {\bibfnamefont {C.~E.}\ \bibnamefont {Carlston}},\ }\href {\doibase 10.1103/PhysRev.140.A1407} {\bibfield  {journal} {\bibinfo  {journal} {Phys. Rev.}\ }\textbf {\bibinfo {volume} {140}},\ \bibinfo {pages} {A1407} (\bibinfo {year} {1965})}\BibitemShut {NoStop}%
\bibitem [{\citenamefont {Medved}, \citenamefont {Mahadevan},\ and\ \citenamefont {Layton}(1963)}]{Medved1963}%
  \BibitemOpen
  \bibfield  {author} {\bibinfo {author} {\bibfnamefont {D.~B.}\ \bibnamefont {Medved}}, \bibinfo {author} {\bibfnamefont {P.}~\bibnamefont {Mahadevan}}, \ and\ \bibinfo {author} {\bibfnamefont {J.~K.}\ \bibnamefont {Layton}},\ }\href {\doibase 10.1103/PhysRev.129.2086} {\bibfield  {journal} {\bibinfo  {journal} {Phys. Rev.}\ }\textbf {\bibinfo {volume} {129}},\ \bibinfo {pages} {2086} (\bibinfo {year} {1963})}\BibitemShut {NoStop}%
\bibitem [{\citenamefont {Ferron}\ \emph {et~al.}(1981)\citenamefont {Ferron}, \citenamefont {Alonso}, \citenamefont {Baragiola},\ and\ \citenamefont {Oliva-Florio}}]{Ferron1981}%
  \BibitemOpen
  \bibfield  {author} {\bibinfo {author} {\bibfnamefont {J.}~\bibnamefont {Ferron}}, \bibinfo {author} {\bibfnamefont {E.~V.}\ \bibnamefont {Alonso}}, \bibinfo {author} {\bibfnamefont {R.~A.}\ \bibnamefont {Baragiola}}, \ and\ \bibinfo {author} {\bibfnamefont {A.}~\bibnamefont {Oliva-Florio}},\ }\href {\doibase 10.1088/0022-3727/14/9/018} {\bibfield  {journal} {\bibinfo  {journal} {J. Phys. D. Appl. Phys.}\ }\textbf {\bibinfo {volume} {14}},\ \bibinfo {pages} {1707} (\bibinfo {year} {1981})}\BibitemShut {NoStop}%
\bibitem [{\citenamefont {Kremers}, \citenamefont {Beijers},\ and\ \citenamefont {Brandenburg}(2013)}]{Kremers2013}%
  \BibitemOpen
  \bibfield  {author} {\bibinfo {author} {\bibfnamefont {H.~R.}\ \bibnamefont {Kremers}}, \bibinfo {author} {\bibfnamefont {J.~P.}\ \bibnamefont {Beijers}}, \ and\ \bibinfo {author} {\bibfnamefont {S.}~\bibnamefont {Brandenburg}},\ }\href {\doibase 10.1063/1.4793375} {\bibfield  {journal} {\bibinfo  {journal} {Rev. Sci. Instrum.}\ }\textbf {\bibinfo {volume} {84}},\ \bibinfo {pages} {025117} (\bibinfo {year} {2013})},\ \Eprint {http://arxiv.org/abs/1301.3287} {arXiv:1301.3287} \BibitemShut {NoStop}%
\end{thebibliography}%

\end{document}